# Anomalous transport in angstrom-sized membranes with exceptional water flow rates and dye/salt rejections

Rathi Aparna [1], Singh Khushwant [1], Saini Lalita [1], Kaushik Suvigya [1], Dhal Biswabhusan [1], Parmar Shivam [2], & Kalon Gopinadhan [1,2,*]

[1] Discipline of Physics, Indian Institute of Technology Gandhinagar, Gujarat 382355, India

[2] Discipline of Materials Engineering, Indian Institute of Technology Gandhinagar, Gujarat 382355, India

(*E-mail: gopinadhan.kalon@iitgn.ac.in)

**ABSTRACT**

**Fluidic channels with physical dimensions approaching molecular sizes are crucial for novel desalination, chemical separation, and sensing technologies. However, fabrication of precisely controlled fluidic channels in the angstrom size is extremely challenging. This, along with our limited understanding of nanofluidic transport, hinders practical applications. Here, we fabricated high-quality salt-intercalated vermiculite membranes with channel sizes ~3-5 Å, highly dependent on intercalant. Unlike pristine samples, the salt-intercalated membranes are highly stable in water. We tested several such membranes, of which 0.6 μm thick membranes showed dye rejection efficiencies >98% with exceptionally high water permeance of 5400 L $m^{-2}$ $h^{-1}$ $bar^{-1}$ at a differential pressure of 0.9 bar. Interestingly, the same membrane also rejected NaCl ions, with efficiencies of ~95%. Our highly confined channels exhibit sub-linear ionic conductance related to hydration sizes, steric exclusion, $K^+$ mobility enhancement, and conductance saturation at concentrations ≤ 10 mM. This makes highly confined channels interesting for both fundamental science and applications.**

1. Introduction

Angstrom-scale channels play a vital role in many essential functions of life; for example, $Na^+$ channels of 3 to 5 Å diameters exhibit a high $Na^+/K^+$ selectivity of 10 to 30 [1]. 3 Å diameter *Aquaporin* protein channels transport water molecules selectively while rejecting most of the ions including protons. The high selectivity of *Aquaporin* channels is related to its hydrophobicity and angstrom sizes [2]. Several groups attempted to mimic biological channels, and had good success in the fabrication of channels/pores with sizes in the nanometer range or smaller. The highly confined channels exhibit fast water flow[3], hydration[4] and size-based selectivity[5], anomalous dielectric constant[6], ion mobility enhancement[7], giant osmotic energy[8,9], etc. All these observations are just indicative of the rich science that exist at the smallest scale.

On the practical side, angstrom-sized channels with sub-micron membrane thickness promise improved filtration performance, low cost and low operating pressures. Current state-of-the-art membranes such as Toray TM 610, NF 270, and Desal 5 L, reject dyes with efficiencies of 98 to 99% and transport water molecules with a flux of 212 L $m^{-2}$ $h^{-1}$ at a pressure of 15 bar [10]. These membranes also reject NaCl with efficiencies of < 50% at differential pressures > 6 bar. Achieving high water flux along with good ion and dye rejection at low operating pressures still remains a challenge. This is partially due to our inability to fully resolve the competing transport mechanisms at the angstrom scale. Inspired by the performance of biological channels, several materials were investigated. Among these, two-dimensional (2D) materials are distinctly different and have atomically small thickness [11]. Notable examples are graphene slits [3,5], graphene oxide laminates [12,13], carbon nanotubes [7,14], BN nanotubes [9], nanopores [15], and clay membranes [16,17]. Most of these 2D material-based membranes, swell in water, resulting in poor sieving characteristics. Several studies tried to address this issue by focusing either on mechanical confinement[18] or strengthening the interaction between the nanosheets[12,19,20]. Although, several of these membranes showed good dye rejections, the water permeance always remained < 1000 L m$^{-2}$ h$^{-1}$ bar$^{-1}$ for sub-micron thick membranes.

To overcome this, we chose crystals of vermiculite, the clay material, for several reasons. Vermiculite crystals are abundant and layered, therefore highly suitable for making membranes consisting of two-dimensional (2D) laminates. The interlayer space in bulk vermiculite crystals has unintentionally intercalated hydrated cations occupying a space of 3-5 Å. This space is similar to or larger than the size of water molecules. Manipulating these spaces with controlled intercalation of salt ions could lead to new generation membranes. Structurally, vermiculite has two tetrahedral and one octahedral sheet made up of $O^{2-}$, in which $Si^{4+}$ ions occupy the tetrahedral sites and $Al^{3+}$ ions the octahedral sites. However, the substitution of $Al^{3+}$ in place of $Si^{4+}$ at about one-quarter of the tetrahedral sites and substitution of $Mg^{2+}$ and $Fe^{2+}$ in place of $Al^{3+}$ in aluminium hydroxide octahedral sheets leaves a net negative layer charge. This is balanced by the intake of various cations like $Ca^{2+}$, and $Mg^{2+}$ in the hydrated form, occupying interlayer spaces of vermiculite[21].

In our study, these unintentionally intercalated cations were first replaced with lithium ions, resulting in expanded vermiculite crystals because of the free swelling nature of $Li^+$ [22]. The expanded vermiculite was exfoliated to construct membranes of 2D laminates. We, however, observed that the Li-vermiculite (Li-V) membrane is not stable in water. To make it water-stable, we treated pristine free-standing Li-V membranes with aqueous solutions containing one of the following ions, $Al^{3+}$, $Ca^{2+}$, $Na^+$, or $K^+$. The salt intercalation indeed helped the membranes to be stable and also tuned the interlayer spaces of vermiculite. These membranes were observed to be highly efficient for salt and dye rejections and also exhibit exceptionally high water flux.

## 2. Experimental section
### 2.1. Membrane fabrication

We fabricated lithium intercalated vermiculite membranes via a two-step ion exchange process. In the first step of the exchange process, 100 mg of vermiculite was heated in 200 ml of saturated sodium chloride solution at 100°C for 24 hours in a reflux assembly, a procedure very similar to other reports [23,24]. The resultant sodium exchanged crystals were repeatedly washed with DI water to remove excess NaCl. These crystals were then heated in 200 ml of 2 M LiCl solution for 24 hours and again washed with DI water to remove excess LiCl that resides on the surfaces. This product was then dried, weighed, and dispersed in DI water with a concentration of 1 mg/ml. The solution containing expanded Li-vermiculite crystals was subjected to ultra-sonication (power = 24 W) in water for 15 minutes. The suspension was then centrifuged at 3000 rpm for 10 minutes to obtain a uniform solution consisting of thinner layers. The supernatant gave a light-yellow dispersion of vermiculite layers (Fig. 1a) and the final solution consisted of 0.62 mg/ml of vermiculite nanosheets giving a 62% yield. Zeta potential measurements provided a value of -35±1 mV, a strong indication of a negative surface charge on the suspended flakes. The successful dispersion of vermiculite layers in water is confirmed by observing the Tyndall effect (Fig. 1a). We prepared a Li-vermiculite membrane, labeled as Li-V, of diameter 1.6 cm via vacuum filtration on several porous supports (pore size = 100-200 nm) such as AAO, and PVDF. The vermiculite naturally peeled off from the porous support in the case of thicker membranes (Top inset of Fig. 1a); however thin membranes were difficult to handle without the porous support. We prepared several membranes of thickness ranging from 0.4 to 5.0 µm. We performed our ion transport measurements with free-standing vermiculite membranes, which helped us to understand their intrinsic properties without the influence of porous support. In pressure-assisted filtration experiments, we used membranes with PVDF porous support as the latter provide mechanical strength and a smooth surface to avoid wear and tear of the membrane.

*2.2. Pressure-driven filtration setup*

For the salt and dye permeation experiments, we used a pressure-driven filtration setup, where the feed side of the membrane was exposed to the atmosphere and the other side to pressures lower than 1 bar. We conducted the permeation test under several pressure gradients from 900 mbar to 50 mbar. The permeate was collected and analyzed to quantify the water permeance and the concentration of dye or ions, if present any.

2.3. *Ion transport setup*

The setup used for ion transport studies consists of a tau cell made of polyether ether ketone (PEEK) material holding a 5 µm thick vermiculite membrane separating salt solutions on either side of the membrane. The membrane was mounted on an acrylic sheet that has a pre-drilled square hole, with the help of epoxy glue (Stycast 2625), which ensures that the only path for ion transport is through the vermiculite membrane. The actual area of the membrane under study was 4 mm$^2$. Two Ag/AgCl electrodes were used to measure the ionic current with the help of a source-meter unit (Keithley-Tektronix 2614B) and LabVIEW software.

**3. Results and Discussion**

The free-standing Li-V membranes are found to disintegrate within 2 - 5 minutes of exposure to both water (Fig. 1b) and aqueous LiCl solutions having a maximum concentration of 0.1 M (Fig. S1). However, Li-exchanged vermiculite membranes appear stable in concentrated LiCl solutions, for example, 1 M and above. If the stable membrane is transferred from 1 M LiCl solution to water, it quickly disintegrates making them unsuitable for water-related applications. To make them water-stable, we immersed Li-V membranes in one of the salt solutions of KCl, NaCl, CaCl$_2$, or AlCl$_3$ of 1M chloride concentration for 24 hours. After the ion exchange, the membranes were thoroughly washed with DI water which removed the residual salt that resides on the sample surface. We performed several measurements to confirm the successful exchange of K$^+$, Na$^+$, Ca$^{2+}$, or Al$^{3+}$ with Li$^+$ in the Li-V membranes. The exchanged membranes are labelled as cation-V membranes. The salt-exchanged membranes were found to be water stable for at least six months, without any evidence of swelling in aqueous solutions (Fig. 1c). In addition, these membranes were also tested under large concentration gradients and maximum applied voltages of ±300 mV across the membrane and were found to be stable (Fig. S2). We did not observe any voltage-induced delamination in any of our salt-stabilized membranes. Though the synthesis of Li-vermiculite membranes were reported previously[23,24], their stability in water was rarely discussed. In those studies, the membrane exposure to water was only for a short duration, and hence the instability might not be very obvious. We also checked the water wettability of these membranes with contact angle measurement and they are found to be mostly hydrophilic (Fig. S3).

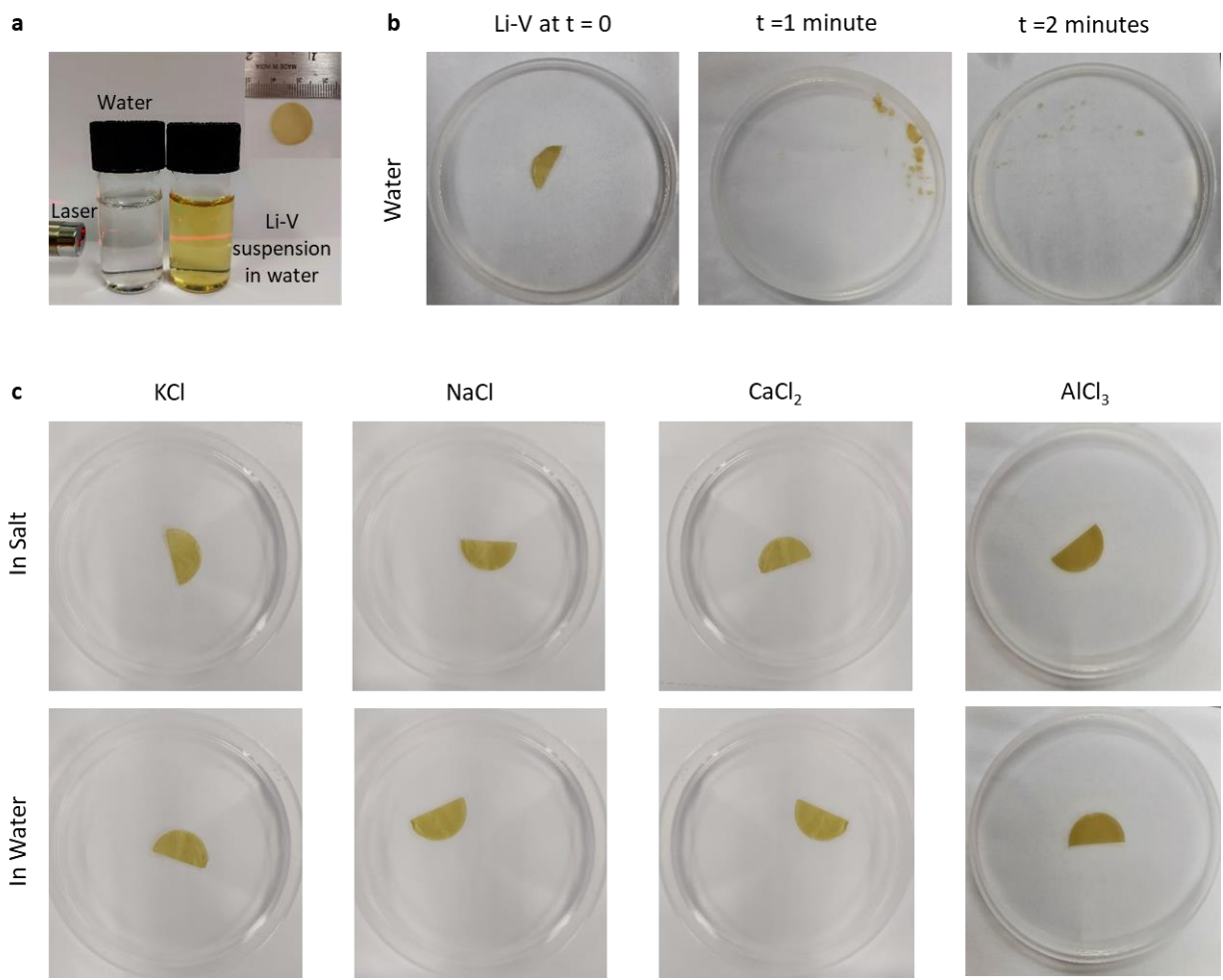

**Fig. 1. Stability of free-standing vermiculite membranes in water.** (a) The exfoliated vermiculite layers are successfully dispersed in water as evident from the characteristic Tyndall effect. The solution has a typical concentration of 0.62 mg/ml. Inset: pristine free-standing Li-V membrane. (b) A free-standing Li-V membrane disintegrates in water within 2 - 5 minutes of exposure. (c) Camera images of a Li-V membrane stabilized in salt solutions of KCl, NaCl, CaCl$_2$, and AlCl$_3$.

Having confirmed the water-stability of our membranes, we investigated the microstructure of our pristine and intercalated membranes. Using the atomic force microscopy (AFM) technique (Fig. 2a), the average flake thickness of exfoliated Li-vermiculite was determined to be ~1.5 nm (Fig. 2b), which corresponds to 1 layer of vermiculite. The cross-section of the membrane obtained using scanning electron microscopy (SEM) shows an exquisite laminate structure (Inset of Fig. 2b), and the surface of the membrane shows a continuous microstructure without any obvious pinholes (Fig. S2). The X-ray diffraction (XRD) data recorded for bulk vermiculite crystal shows multiple peaks around 8.65°, indicating the presence of numerous unintentional cations in the interlayer space (Fig. S4). The XRD data of the Li-V membrane (Fig. 2c-d) shows a sharp, and intense peak at 2θ = 7.05°, corresponding to (001) plane. The peak at 7.05° (Full width at half maximum= 0.81°) provides an interlayer separation, *d* of 12.5 Å, confirming that the space between the layers has been successfully exchanged with Li -ions (Fig. S4). We also recorded the XRD data of salt-stabilized membranes (Fig. 2c-d). In Na-V, Ca-V, and Al-V, apart from the peak (001), we observed higher-order peaks (00$l$) with $l$= 2,3,4 and 5, and a smaller full width at half maximum when compared to Li-V membranes. This indicates that the salt-stabilized laminates are of the highest crystalline quality, homogeneous, and without inter-stratification. The most intense peak (001) is left-shifted from 7.05° to 5.95 ± 0.05° for $Na^+$, $Ca^{2+}$ and $Al^{3+}$ cations, which is related to the modified layer charges arising from the exchange of cations, $Na^+$ is an exception due to its robust hydration shell as compared to $K^+$[25]. Moreover, water layer can either be 1 or 2 and not in between, explaining the sudden change in the interlayer space from ~12 Å to ~15 Å. Unlike other salt-stabilized membranes, the XRD of K-V membranes show only (001) and (004) peaks, which is very similar to the XRD pattern of Li-V membranes. When compared with Li-V, the (001) peak of $K^+$ is shifted very little to the right from 7.05° to 7.25°. The little difference in the structure of Li-V and K-V could be related to the similar layer charges arising from the exchange of $Li^+$ with $K^+$.

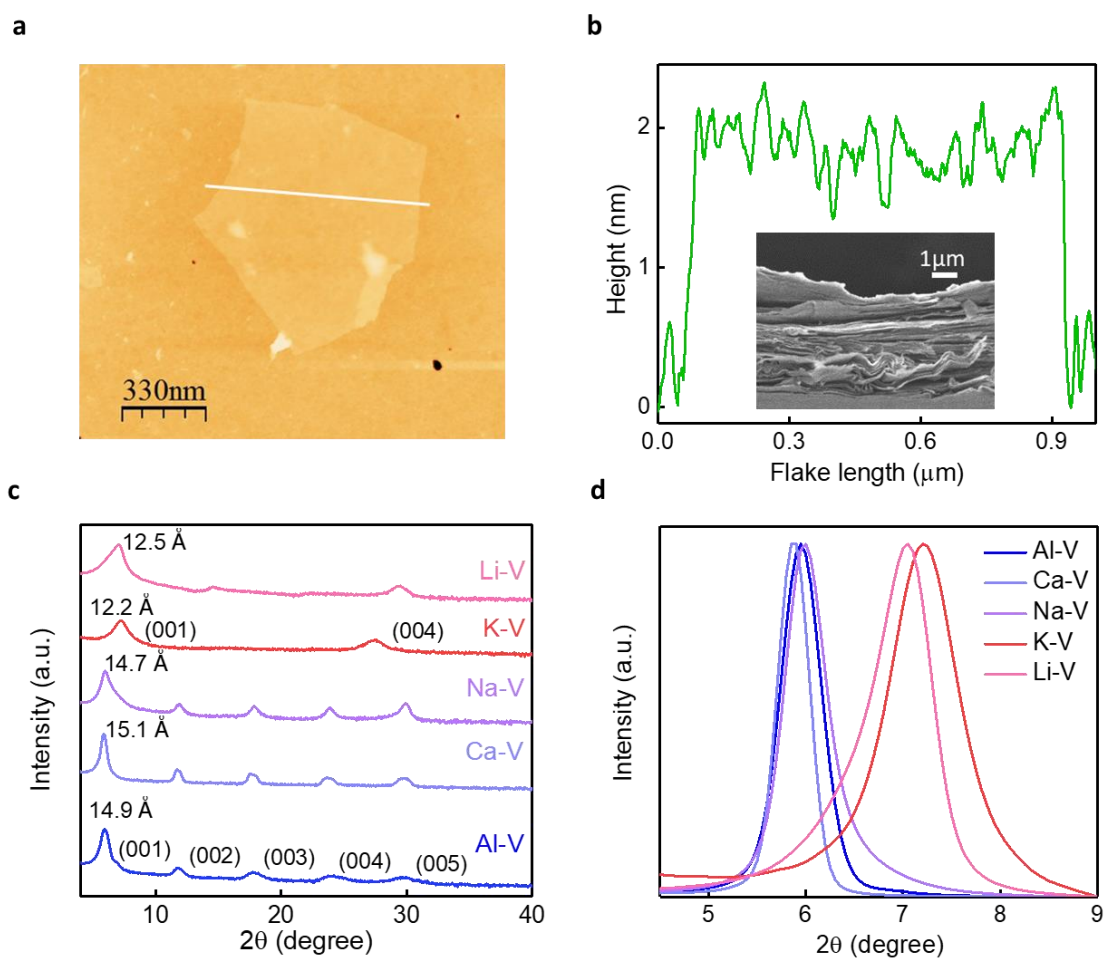

**Fig. 2. Characterization of vermiculite membranes.** (a) AFM image of an isolated vermiculite flake. (b) The height profile of the flake estimated a thickness of ~1.5 nm. Inset: the cross-sectional SEM image of the membrane. (c) X-ray diffraction data of Li-V, K-V, Na-V, Ca-V, and Al-V membranes. Higher-order peaks are clearly seen in most of the cation exchanged V-membranes. (d) Comparison of the intense peak (001) of Li-V, K-V, Na-V, Ca-V, and Al-V membranes. (001) peak of Li-V shows an appreciable shift upon Na, Ca and Al- exchange, however, with K, it is not affected much.

Given the immense interest in dye removal from industrial wastewater and the high quality of our membranes, we investigated the water/dye permeation properties of Na-stabilized membranes. We tested our membranes with several dyes such as methyl orange (MO, anionic), crystal violet (CV, cationic), rhodamine 6G (R6G, cationic), methyl blue (MB, anionic), and brilliant blue (BB, anionic)[16]. The largest dimension of the dye molecule that we tested ranges from 1.19 nm to 2.73 nm. The schematic of the filtration setup is shown in Fig. S5a. The dye concentration was approximately 10 mg/L on the feed side. We collected the retentate, feed, and permeate solutions and analyzed them with the help of a UV-Visible spectrometer (Fig. S5b-f). The UV-Visible spectra of retentate solution show a broad and intense peak in the visible region for all the studied dyes. In contrast, the spectral intensity of the permeate solution, was extremely weak or within the detection limit of the instrument. Since intensity is a measure of the concentration, it is clear that the dye molecules on the permeate side are significantly low. The rejection, R was calculated by recording the relative difference in the concentration of permeate ($C_p$) and feed ($C_f$) compartments as

$$R (\%) = (1-C_p/C_f) \times 100\%$$

Our membrane shows rejection of >98 % for MO, >99 % for CV, >99 % for R6G, >99 % for MB and >99 % for BB (Fig. S5b-f). The water permeance was estimated from the collected volume, V, the pressure difference, ΔP, the duration of water permeation, t, and the membrane area, A. The permeance, J, is given by

$$J = V/(\Delta P*A*t)$$

We recorded the highest permeance of ~7183 L $m^{-2}$ $h^{-1}$ $bar^{-1}$ for pure water with a ~ 400 nm thick Na-V membrane; however, it reduced to 6600 L $m^{-2}$ $h^{-1}$ $bar^{-1}$ for the brilliant blue dye with a rejection of ~90%. When we increased the membrane thickness to ~ 600 nm, the rejection increased to > 99% with a slightly reduced water permeance of 5400 L $m^{-2}$ $h^{-1}$ $bar^{-1}$. For thicker membranes beyond 600 nm (Fig. S6), the rejection remained >99%; however, the water permeance scaled inversely with the thickness (Fig. 3a). Since our measurements were static, the dyes accumulated on membrane surface over time, reducing the water permeance, as shown in Fig. S7a. The dynamic measurements could help minimizing the dye accumulation. We also note that both anionic and cationic dyes did not show any significant difference in rejection (Fig. S7b), suggesting the dominance of size-based rejection. We repeated these measurements with other salt-stabilized membranes also and observed that with increase in interlayer cation valence, the water permeance decreased by a factor of 2.5, with negligible change in the rejection (Fig. S7c). The reduced water permeance is inferred to be a result of decreased hydrophilicity and hence a weaker intake of water.

Removal of salt from seawater is extremely important for potable water applications, and we explored this possibility with our 1.2 µm thick Na-stabilized membranes. We tested the permeation of NaCl (1 M) ions with a pressure-assisted filtration setup at several pressure gradients, up to a maximum of 900 mbar. We quantified the permeated concentration using inductively coupled plasma-optical emission spectroscopy (ICP-OES). At ΔP of 50 mbar, we observed the highest rejection of ~95% and a water permeance of ~120 L m$^{-2}$ h$^{-1}$ bar$^{-1}$ (Fig. 3b). When ΔP = 550 mbar, the ion rejection decreased to ~8% and the water permeance increased to > 2300 L m$^{-2}$ h$^{-1}$ bar$^{-1}$ (Fig. S7d). When we further increased ΔP to 900 mbar, a rejection of > 7% was recorded. We repeated these measurements with several membrane thicknesses at ΔP = 50 mbar. The permeate concentration increased from ~50 to 650 mM upon decrease in thickness from 1.2 to 0.4 µm (Fig. 3b).

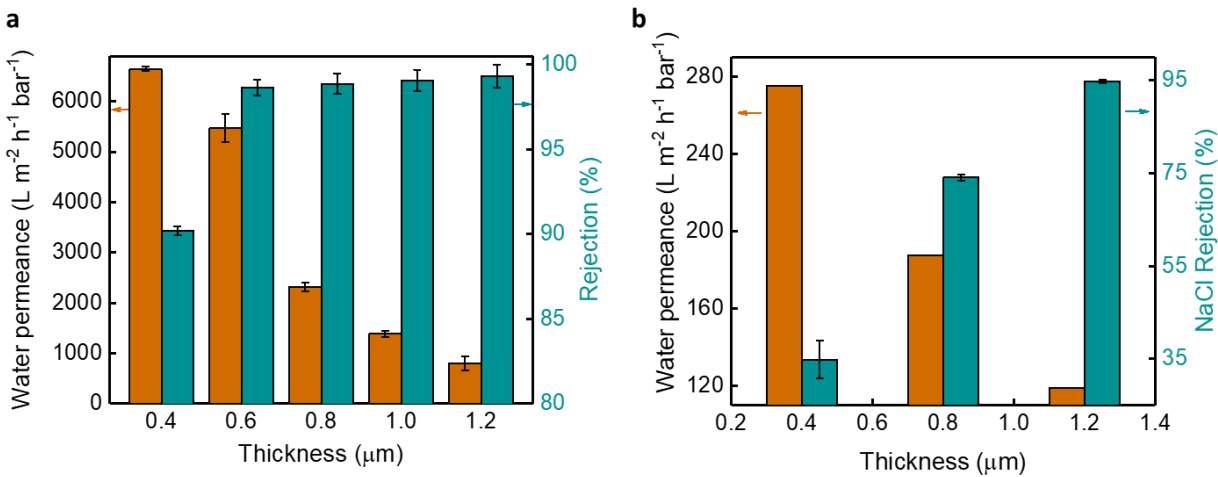

**Fig. 3. Dye and Salt filtration.** (a) Water permeance (left Y-axis) and Brilliant blue dye rejection (Right Y-axis) as a function of membrane thickness at ΔP = 900 mbar. (b) Water permeance (Left Y-axis) and NaCl rejection (Right Y-axis) as a function of membrane thickness at ΔP = 50 mbar. The pressure gradient clearly changes the membrane permeability, as evident from the lower water flow rates at ΔP = 50 mbar as compared to ΔP = 900 mbar. This is inferred to be a result of smaller interlayer distance at lower pressure gradients. The error bar is for 3 samples.

We also performed these experiments with a mixture of 25 ml 1 M NaCl and 25 ml BB dye (concentration ~ 10 mg/L), which resulted in a final 50 ml solution of 0.5 M NaCl and 5 mg/L BB. Using a 1.2 µm thick Na-V membrane at ΔP = 250 mbar, we were able to get a salt rejection of ~ 40% and a dye rejection of ~ 99% with water permeance of 1290 L m$^{-2}$ h$^{-1}$ bar$^{-1}$ comparable to that of permeation experiments with only salt or dye. But at a higher differential pressure of ΔP = 550 mbar, the salt rejection decreased to ~ 10% while the dye rejection remained constant with increased water permeance of 2277 L m$^{-2}$ h$^{-1}$ bar$^{-1}$, the flux was comparable to only 1 M NaCl as the feed (Fig. S7d). This confirms the ability of our membranes to separate salt and dye from a mixed solution (Fig. S8).

In addition, we examined the ion permeation across 5 μm thick free-standing membranes with a forward osmosis experiment. For this, we kept 1 M NaCl solution on the feed side and DI water on the permeate side (Fig. S9a). During a period of ~45 hours, sufficient for establishing equilibrium, a concentration of ~10 mM of ions permeated through the membrane. This also agrees with the result of UV-visible spectra (Fig. S9b). In this case, the ion rejection was estimated to be 99%. The higher salt rejection is due to the larger thickness of 5 μm for the membrane used in this experiment. It is to mention that due to osmosis, there is a flow of water from permeate to feed side; hence, our measured ion concentration is in the upper limit.

We further examined the little permeation of salt ions across 5 μm thick membranes to understand the rejection mechanism. The approximate size of the fluidic channel that is responsible for the transport is estimated by considering the thickness of silicate layers (~9.6 Å)[26] and the interlayer distance, $d$, where the latter is taken from the XRD data. In the case of K-V membranes, $d_{(001)} \approx 12.2$ Å, leaving a fluidic space that can accommodate ~1 layer of water molecules. However, in the case of Na-V, Ca-V, and Al-V membranes, $d_{(001)} \approx 15$ Å, suggests an interlayer space that can accommodate ~2-layers of water molecules. We performed transport measurements with an equal salt concentration on both sides of the membrane. The hydrated diameter of the cations considered here ranges from 6.6 to 9.6 Å [27]. The hydrated diameter of Cl$^-$ is 6.6 Å [27]. The schematic used for the measurement is shown in Fig. S10a. We first measured KCl ionic conductance through the K-V membrane. To transport the ions, a voltage, $V$, was applied across the membrane, and the resulting current, $I$, was monitored. The measured $I$-$V$ characteristics (Fig. 4a) at various KCl concentrations, $C$, is linear within our maximum applied voltage of ± 300 mV. The conductance, $G$=I/V, is found from the slope of the curves. The measured pH of our deionized (DI) water was ~5.5; therefore, chloride concentrations of 10$^{-5}$ M and 10$^{-6}$ M would roughly correspond to that of DI water. The measured water conductance of twelve samples is plotted in Fig. S10b. We observed a very different ionic conductance behavior for all the salt-stabilized membranes compared to other nanochannel systems. In the concentration range 10$^{-5}$ M - 10$^{-2}$ M, the conductance increased sub-linearly with concentration (Fig. 4b). The measurements with NaCl, CaCl$_2$, and AlCl$_3$ solutions using respective salt-stabilized membranes, also showed similar sub-linear $G(C)$; however, the magnitude of the conductance varies with salt ion. For all the salts, $G \propto C^\alpha$, with the exponent, α = 0.67 to 0.24, inversely related to the hydrated diameter of the cations in the order K$^+$< Ca$^{2+}$< Al$^{3+}$. This points to the role of steric hindrance and ion-ion interactions in providing a size-based discrimination. A similar variation, $G(C)$, was reported in carbon nanotubes[7], and biological channels[28], where the exponent was less than 1. For concentrations ≥ 10$^{-2}$ M, the ionic conductance of all the membranes shows saturation (Fig. 4b). When compared to the bulk conductance of similar geometry, the NaCl conductance through Na-V is observed to be smaller by at least 300. This yields a salt rejection of > 99%, which agrees with the results of ICP-OES and UV-visible spectra (Fig. S9).

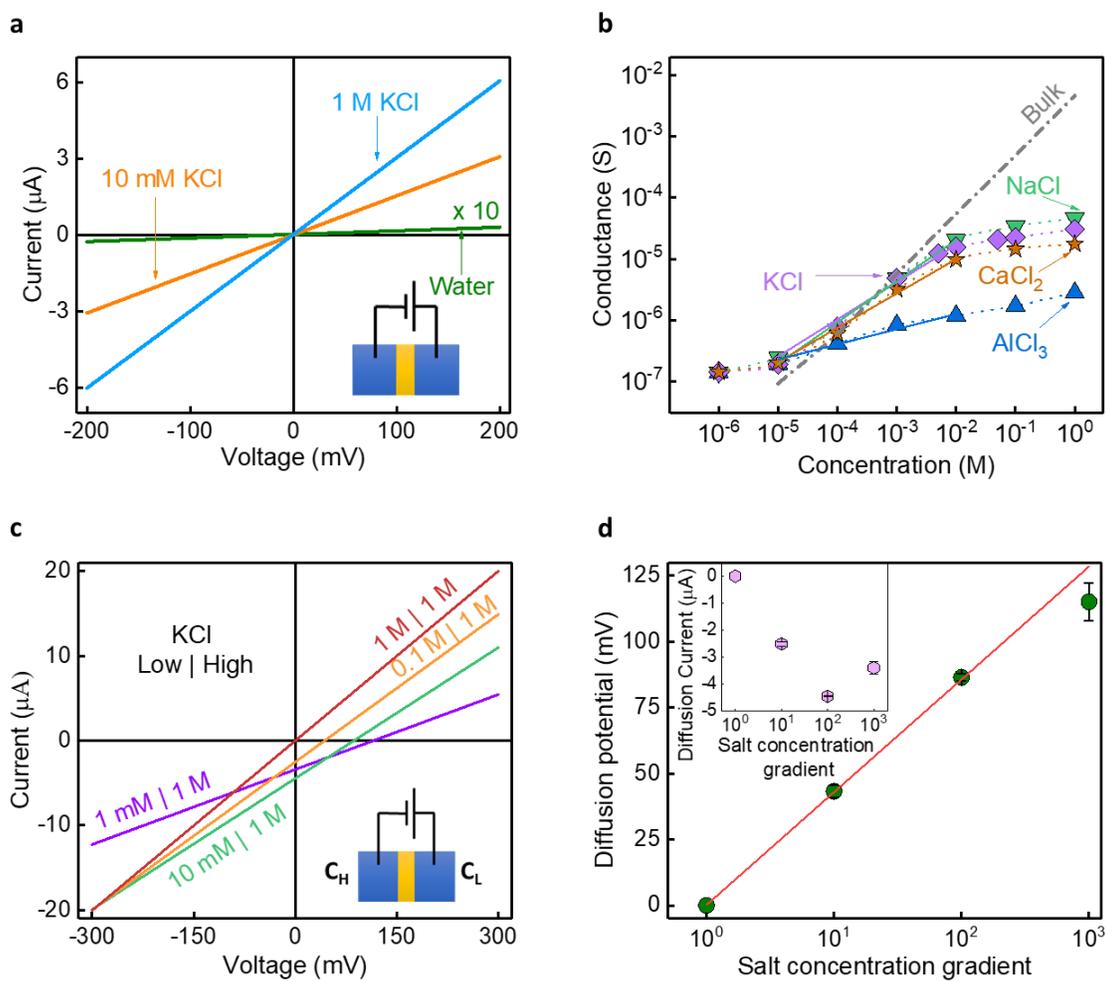

**Fig. 4. Ion transport through salt-stabilized vermiculite membranes.** (a) *I-V* characteristics of K-stabilized vermiculite membrane for several concentrations of KCl. Inset: Schematic of our ion transport measurement setup. (b) Ionic conductance of salt-stabilized membranes with their corresponding salt solutions. For example, K-V membrane is measured with KCl solution. The grey dotted line is the data for open-hole conductance (membrane dimensions are considered) of NaCl calculated from the bulk conductivity reported in literature[29]. Dotted lines are guide for the eye, while the solid lines are fitted with $G \propto C^\alpha$, where α lies between 0.67 and 0.24. (c) *I-V* curves for 1 M KCl on one side of the K-V membrane and 1 mM to 1 M on the other side after subtracting the redox potential. (d) The variation of diffusion potential and current (Inset) as a function of KCl gradient. The diffusion potential is fitted using the Nernst equation, which is shown as a solid red line. The diffusion potential and current are plotted after subtracting the contribution of redox potential at the electrodes.

It should be noted that interlayer cations of vermiculite membranes are exchangeable if the membrane is kept in a solution of higher cation concentration than the one/s already present at the interlayer sites. For example, we kept K-V membrane in AlCl$_3$ solution of concentration greater than 0.1 mM and for atleast 24 hours, which converted K-V into Al-V. The ion exchange is further verified using techniques such as XRD, contact angle measurements and estimating the ionic conductance of the membrane. We monitored the ionic conductance of K-V membrane when kept in 1 M $AlCl_3$ solution where the conductance decreased continuously until it became stable and close to the ionic conductance of 1 M $AlCl_3$ with Al-V (Fig. S4b).

Salt-stabilized vermiculite laminates have different layer charges which could influence the ion transport. To understand this, we performed electro-diffusion measurements with different concentrations of KCl on opposite sides of the membrane (Fig. 4c). The schematic of the measurement setup is shown in the inset of Fig. 4c. We have varied the concentration gradient ($\Delta$) from 10 to 1000. In the absence of any applied voltage, a negative current is measured, and its magnitude increases with increase in $\Delta$ up to 100. Considering the polarity of the electrodes used in the study (Inset of Fig. 4c), the observed negative current (Inset Fig. 4d) suggests that these membranes preferentially transport cations over anions. The applied voltage that is required to nullify this current is designated as zero-current potential, $V_o$. We subtracted the redox potential, $V_R$ from $V_o$, to estimate the net diffusion potential, $V_{diff}$, of the membrane for a particular concentration gradient. The diffusion potential is observed to follow a logarithmic dependence with concentration gradient. This allowed us to estimate the selectivity, S, of the membrane using the Nernst equation (1) as,

$$V_{diff} = S\frac{k_BT}{e}ln\Delta \quad \ldots\ldots\ldots\ldots\ldots\ldots (1)$$

Where $S$ is the selectivity of the membrane, $k_B$ is the Boltzmann constant, $T$ is the temperature (298 K), and $e$ is the elementary charge. For an ideal cation-selective membrane, $S = 1$ and for a non-selective membrane, $S = 0$. We fitted the variation of diffusion potential with KCl concentration gradient (Fig. 4d) using Eqn.1, providing a selectivity of ~0.67. This suggests that K-stabilized vermiculite membranes are cation-selective and the diffusion measurements hint the contribution of surface charge in the ion transport. We measured other salt-stabilized membranes also, and most of them are found to be cation-selective, with almost no selectivity for $AlCl_3$. The selectivity is in the order KCl ≈ NaCl > $CaCl_2$ > $AlCl_3$ (Fig. S11a-b).

We estimated the relative mobility of cations to anions, $\mu_+/\mu_-$, from the measured diffusion potential (Fig. 5a) with the help of Henderson equation[30] as,

$$\frac{\mu_+}{\mu_-} = -\frac{z_+}{z_-}\frac{\ln(\Delta) - z_-FV_{diff}/RT}{\ln(\Delta) - z_+FV_{diff}/RT} \quad \ldots\ldots\ldots\ldots (2)$$

Where $z_+$ and $z_-$ are the cation and anion valances, respectively; F is Faraday's constant; R is the universal gas constant; and T= 298 K. The mobility ratio (cations to anions) is in the order KCl ≈ NaCl > $CaCl_2$ > $AlCl_3$. With the assumption of constant mobility for chloride ions, the enhanced mobility of $K^+$ is interpreted to be a combined effect of confinement and electrostatic interactions (Fig. 5b). A previous report on graphene-based two-dimensional channels[31] indicate that confined channels of height ~6.6 Å, slightly increase the mobility of $K^+$.

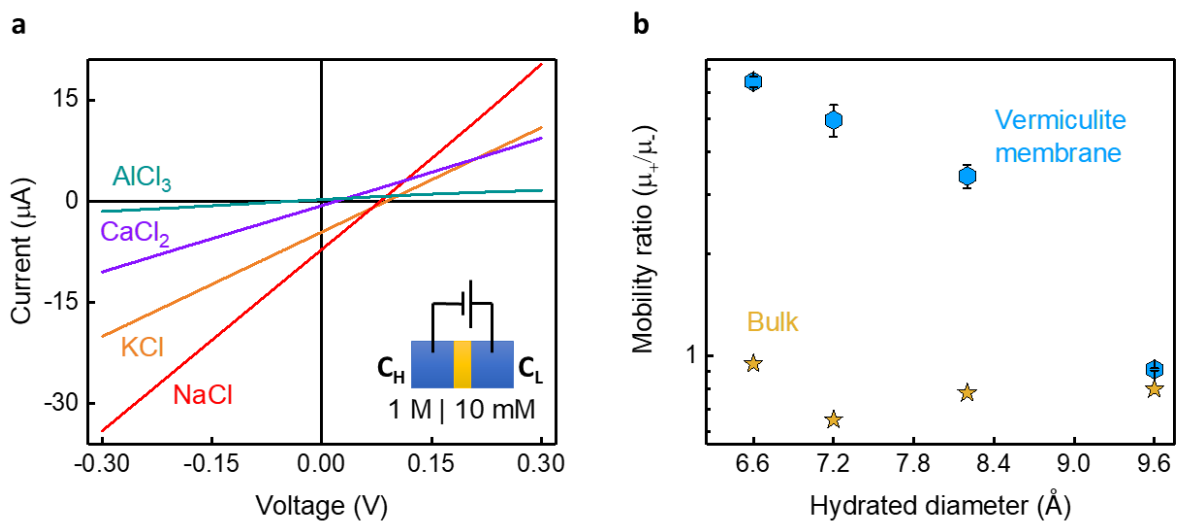

**Fig. 5. Mobility of ions through salt-stabilized membranes.** (a) I-V characteristics of various chloride salts through respective salt stabilized vermiculite membranes with a concentration gradient of 100 across the membrane. (b) Mobility ratio as a function of increasing cation hydration radius through our membranes, which is compared with literature values on bulk systems[29]. Error bar is taken from mobility ratio of the same sample when concentration gradient is 10 and 100.

In clay minerals, the van der Waals attraction of the layers increases as the layer charge increases; an example is the highly compact structure of mica, where the layer charge is 1 with a 2: 1 layered structure. In vermiculite, layer charge is smaller than 1 due to the inclusion of $Ca^{2+}$ and $Mg^{2+}$ in place of $K^+$ in mica. In the case of Al-V membranes, exchanging $Al^{3+}$ with $Li^+$ would reduce the layer charge, explaining the lower selectivity (Fig. S11b). Exchanging the layer with monovalent ions maximizes the layer charge and hence the highest selectivity. In the case of K-V, layer charge is high and hence the van der Waals attraction leads to smaller interlayer spaces, whereas in the case of Al-V, the layer charge is less and hence the interlayer space is large, which agrees with our XRD results. A recent study discusses the possibility of dynamical change in the laminar structure whenever a cation is exchanged. The local delamination removes the existing cation, and subsequently, re-stacking happens, which accommodates the new cation [32]. In the case of monovalent cations, such as $K^+$ and $Na^+$, the larger interlayer spacing with $Na^+$ could be a result of the larger and stable hydration shell of $Na^+$ over $K^+$.

Though diffusion measurements shed light on the layer charge, our many other observations suggest that the incoming species' molecular/hydrated ionic size controls the transport. The pressure-assisted filtration experiments provide insights into the membrane microstructure and its modifications under pressure gradients. We fabricated the membrane by assembling flake sizes of ~600 nm, where the flake thickness is ~1.5 nm. Fig. S7d summarizes the water permeance and NaCl rejection efficiencies as a function of pressure gradients from 50 to 550 mbar. When the pressure gradient is low, the water permeance is low, however, the NaCl rejection is high. When we increased the pressure gradient to 550 mbar, there is less rejection of NaCl; however, the water permeance increased. In the absence or low values of pressure gradient, the flake placements are perfect, and there are only entry/exit points for the water, not the salt. However, when we gradually increased the pressure across the membrane, the interlayer spacing of the laminates increased, allowing the entry/exit of water molecules while rejecting most ions/dyes of sizes larger than the interlayer spacing. Since the interlayer space is small at low pressure gradients, the water permeance will also be low. However, larger interlayer spacing was induced when we increased the pressure gradient to 550 mbar, allowing salt ion permeation and increased water permeance (Fig. S12a-b). Here the interlayer spacing is still in sub-nm, much smaller than the sizes of dye molecules; hence, they are rejected. With the increase in thickness, the water permeance is found to decrease. As thickness increases, due to the corrugated path, the effective travel length of water molecules becomes larger, resulting in reduced water permeance. A very recent report on graphene oxide predict a similar increase in interlayer space under increased differential pressures[33].

The measured water permeance of 5400 L m$^{-2}$ h$^{-1}$ bar$^{-1}$ far exceeds previous reports. There were some studies on hydrophilic membranes of vermiculite[17], montmorillonite[16], and Ti$_3$C$_2$T$_x$[19]; however, they exhibit moderate water fluxes. We measured the water contact angle of the salt-stabilized membranes and found them to increase in the order Na-V < Ca-V < Al-V. The higher water permeance in the case of Na-V membranes is related to the strong hydrophilicity of thin membranes and suggests the importance of capillary pressure. Further, our salt-stabilized membranes can handle pressures of 1 bar at least. Our vacuum filtration system uses a pressure gradient of 1 bar to fabricate the membranes, and it is observed that any external pressure gradients within this limit does not degrade the membrane performance.

Even at a thickness of 600 nm, our membranes showed better rejection while maintaining a high flux compared to other membranes. Brilliant blue, one of the largest dye molecules, showed exceptionally high flux at a small pressure gradient of 900 mbar. We expect that slightly larger pressure gradients would further enhance the flux. The same membrane can also be utilized for desalination applications at low pressures as it exhibits 95% salt rejection and a water flux higher than or comparable to other membranes[16,17]. This result can be further improved by using several micrometer thick membranes and larger pressure gradients. Overall, our inch-size membranes show great promise and are highly suitable for industrial applications. We tested several samples and more than 90% of the membranes showed high water flux, and good rejections.

## 4. Conclusion

In conclusion, we addressed the water instability of Li-V membranes by intercalating them with various chloride salt solutions. The stabilized membranes have highly confined laminates and tunable interlayer spacing and exhibit steric hindrance. These membranes exhibit high salt and dye rejections with water flow rates exceeding the current state-of-the-art membranes. This is mainly due to the ability to tune the transport parameters, which highly depend on external pressure gradients, layer charge, and hydration diameter of transporting molecules/ions. The ability to fabricate water-stable vermiculite membranes will initiate decades of further research on clay-based membranes.

**ACKNOWLEDGMENTS**

This work was mainly funded by Science and Engineering Research Board (SERB), Government of India, through grant CRG/2019/002702 and partially supported by MHRD STARS with grant no. MoE-STARS/STARS-1/405. A.R. acknowledges the Sabarmati fellowship from IIT Gandhinagar. A. R also acknowledges the PMRF fellowship from Ministry of Education, Government of India. G.K. acknowledges the research fellowship from IIT Gandhinagar. We acknowledge the contribution of IITGN central instrumentation facility. We thank Kalyan Raidongia and Raj Kumar Gogoi of IIT Guwahati for very useful discussions.

**DATE AVAILABILITY STATEMENT**

Data is available from the authors upon reasonable request.

# SUPPLEMENTRY INFORMATION

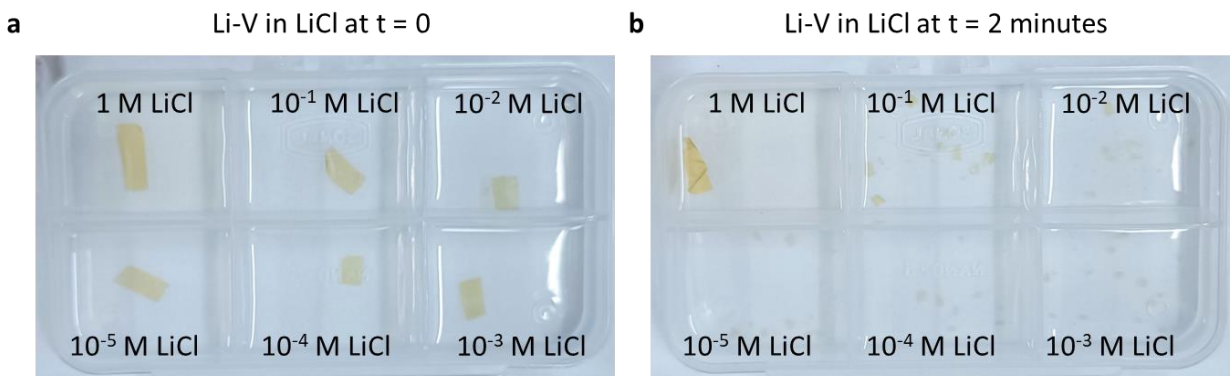

**Figure S1. The stability of Li-V membrane in aqueous LiCl solution.** Li-V membrane in various concentrations of LiCl ranging from 1 M to $10^{-5}$ M at time (a) t = 0, and (b) t = 2 minutes.

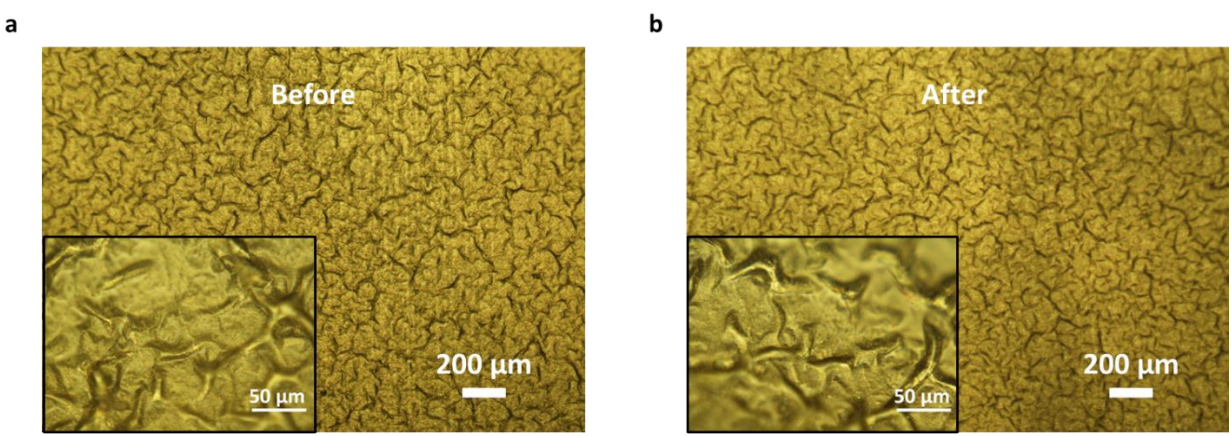

**Figure S2. Surface optical microscope images of Na-V membrane.** Optical images of a Na-V membrane surface (a) before it has been exposed to any voltage and (b) after it has been exposed to a concentration gradient of $10^5$ and a voltage of 300 mV. The inset shows a magnified version of the same surface.

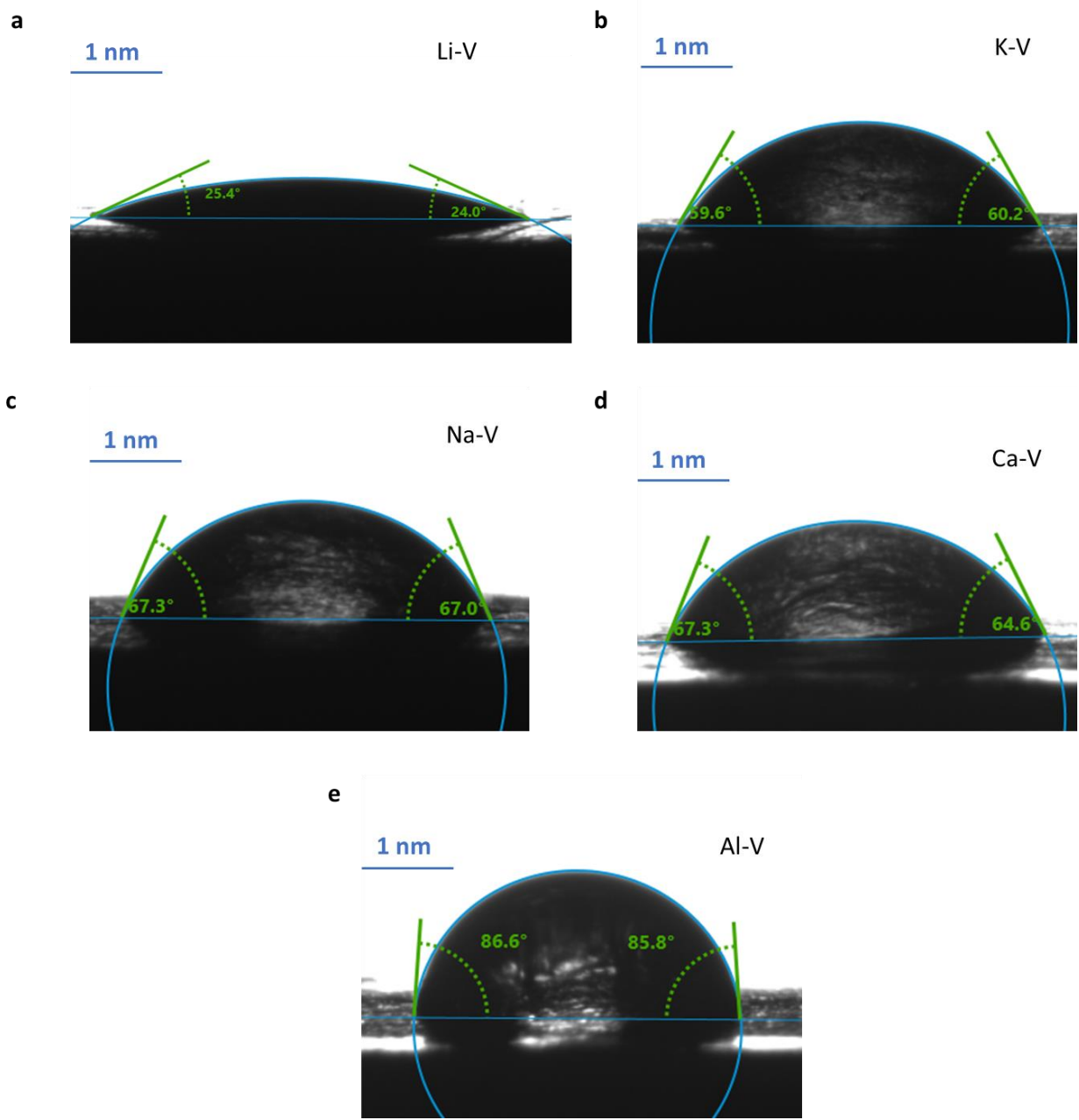

**Figure S3. The hydrophilicity of membranes.** The water contact angle of (a) Li-V membrane, (b) K-V membrane, (c) Na-V membrane, (d) Ca-V membrane, (e) Al-V membrane. A contact angle of < 90º suggests hydrophilicity of all the membranes. As we go from $K^+$ to $Al^{3+}$, the net charge on the layers decreases since vermiculite would require more $Al^{3+}$ to neutralize vacancies. If the layers are charge neutral, then they are expected to be hydrophobic.

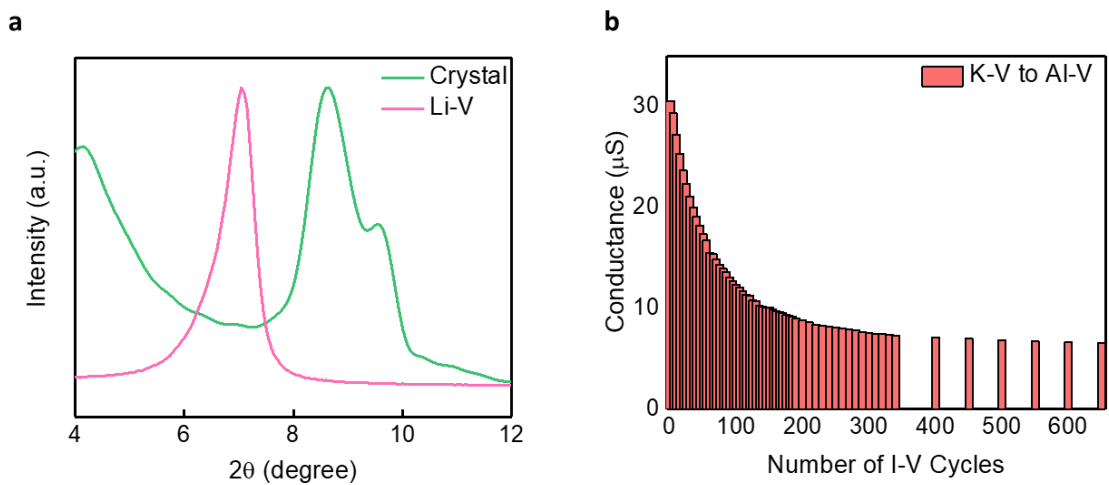

**Figure S4. Additional XRD data and ion exchange in membrane.** (a) XRD peak (001) of raw vermiculite and Li-V membrane, where the former clearly shows multiple peaks, indicating the presence of several cations. (b) Change in the ionic conductance of K-V membrane when it is kept in 1 M AlCl$_3$ solution, which indicates the conversion of K-V to Al-V, and each I-V cycle took almost 2 minutes to complete.

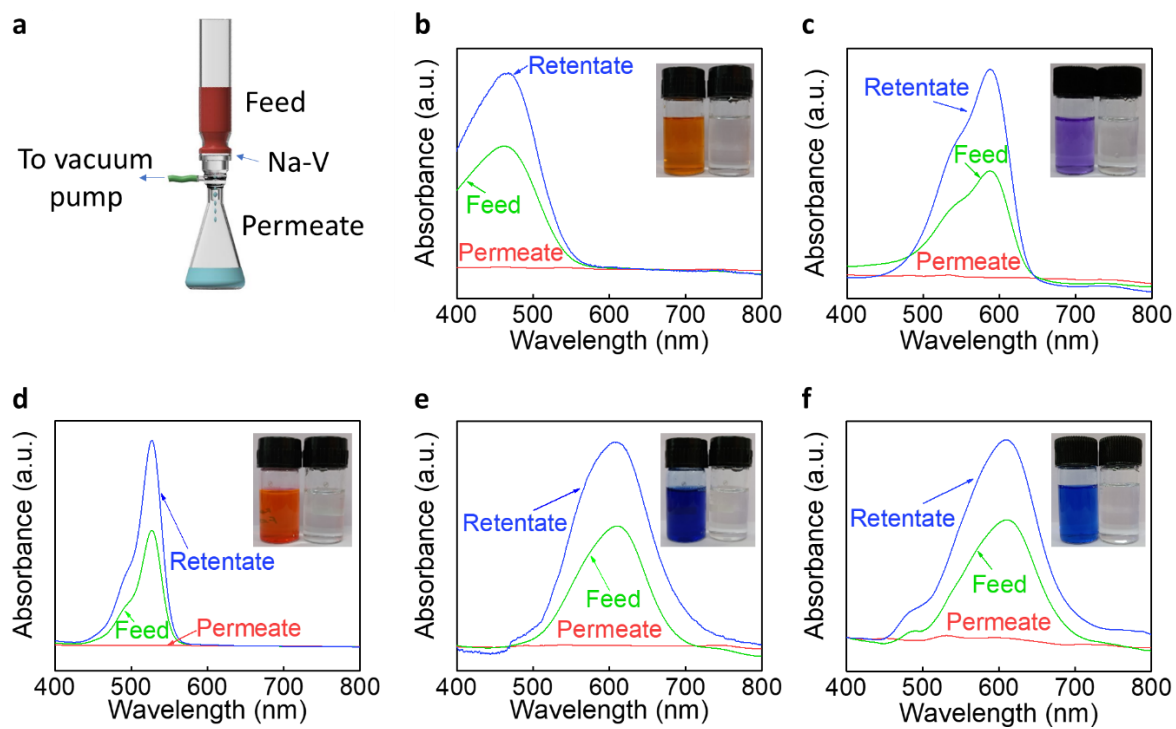

**Figure S5. UV-Vis absorption spectra of dyes.** (a) Schematic of dye filtration set-up. UV-Vis absorption spectra of (b) methyl orange, (c) crystal violet, (d) rhodamine 6G, (e) methyl blue, and (f) brilliant blue. Inset: Feed and permeate solutions. The increase in retentate intensity is an indication that the concentration of the dye is increasing on the feed side due to the outflow of water from feed to permeate.

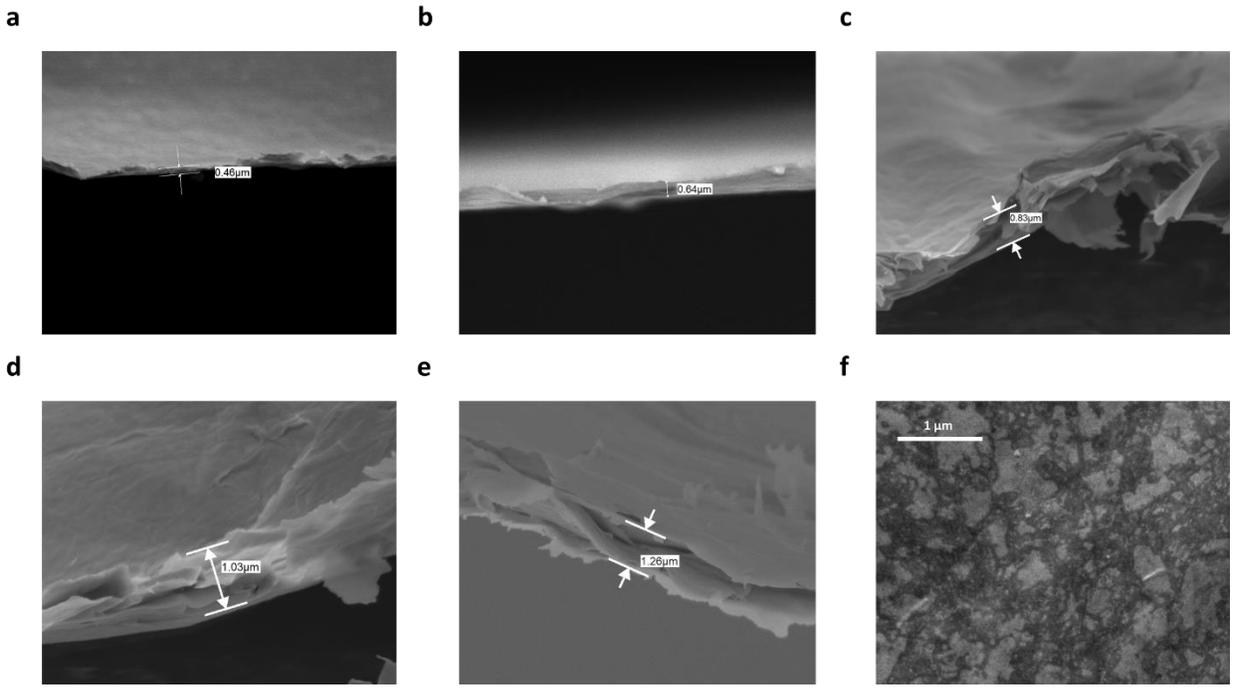

**Figure S6. SEM images of the vermiculite sample.** (a-e) cross-sectional images of Na-V membranes with thickness ranging from ~0.4 µm to ~1.2 µm used for the filtration, and (f) SEM image of Na-V membrane.

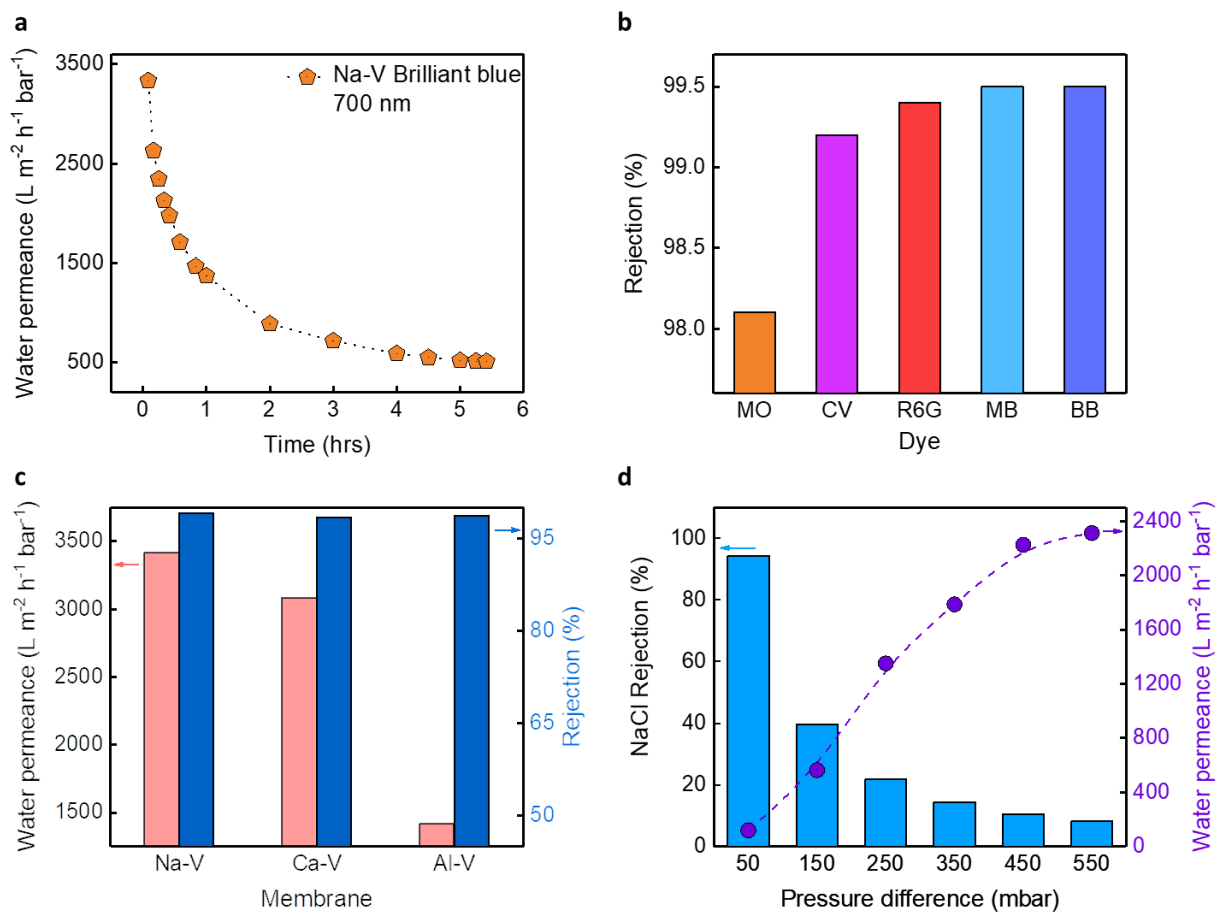

**Figure S7. Rejection and permeance of different salt-stabilized membranes and dyes.** (a) Time dependent water permeance recorded during the filtration of BB dye without any peristaltic pump. (b) Rejection (%) of several dyes with their largest diameter varying from 1.19 nm to 2.73 nm with a ~700 nm thick Na-V membrane where the permeance of water for all the dyes remained ~3200 ± 350 L m$^{-2}$ h$^{-1}$ bar$^{-1}$. (c) Initial water permeance (left Y-axis) and rejection of BB dye (right Y-axis) across ~700 nm thick salt-stabilized membranes. The reduced water permeance of Al-V membranes compared to K-V might be due to less hydrophilicity of Al-V (Fig. S3). (d) Decrease in NaCl rejection (left Y-axis) and increase in water permeance (right Y-axis) with increase in applied pressure gradient through a ~1.2 μm Na-V membrane.

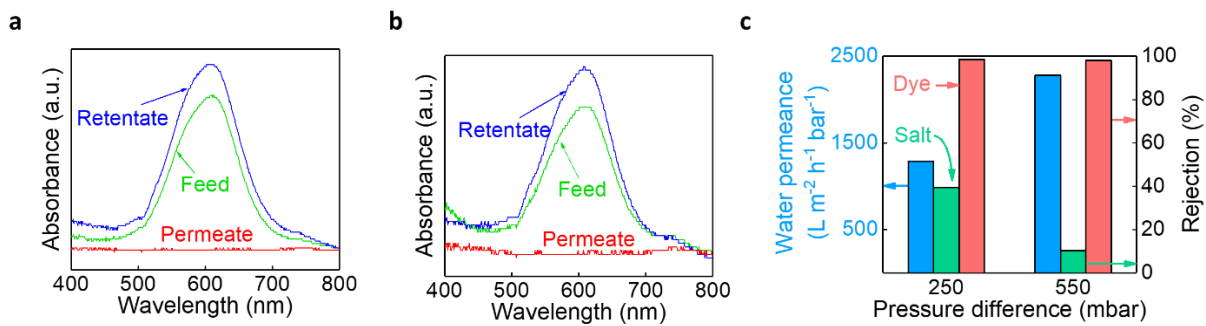

**Figure S8. Dye and salt rejection study from a mixed solution.** (a) At a differential pressure of 250 mbar, (b) At a differential pressure of 550 mbar. Dye rejection remained > 98%. (c) Salt rejection and water permeance at ΔP = 250 mbar and ΔP = 550 mbar through a ~ 1.2 µm thick Na-V membrane.

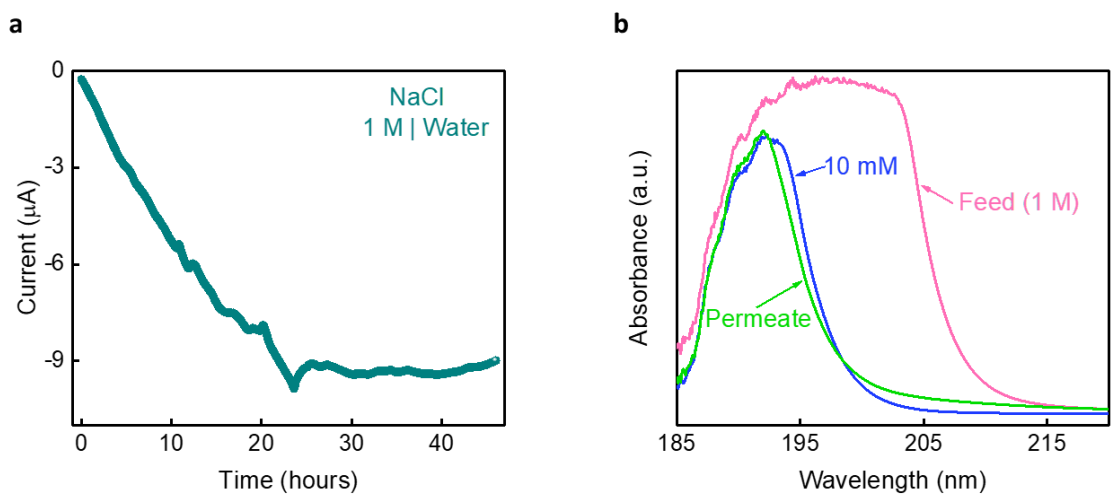

**Figure S9. Forward osmosis experiment.** (a) Diffusion current as a function of time due to the movement of ions from higher to lower concentration. (b) UV-Visible spectra of feed (1 M NaCl), permeate, and the spectra for a 10 mM NaCl control solution, which shows that the concentration of NaCl in the permeate is ~10 mM.

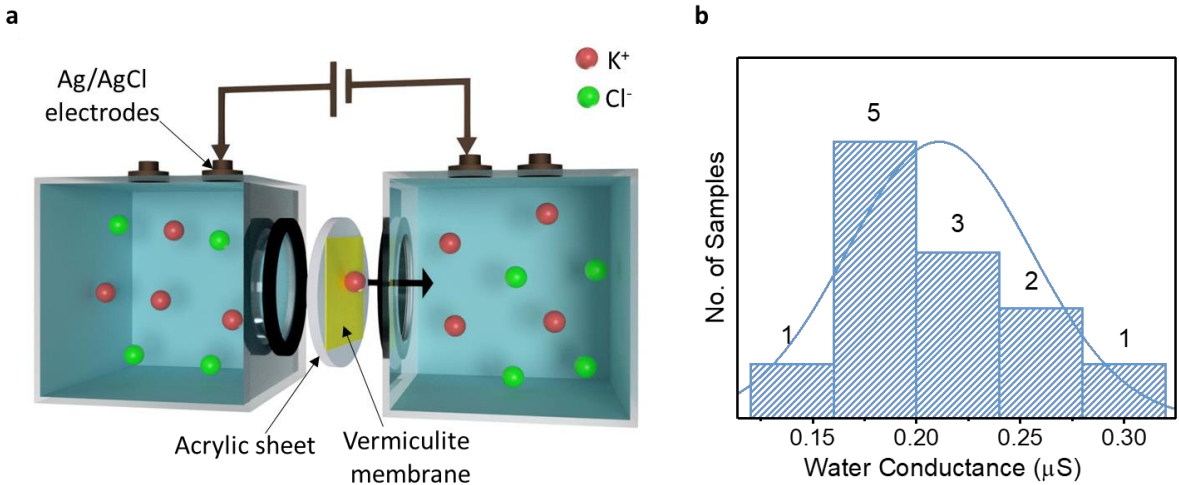

**Figure S10. Ion transport measurement.** (a) Schematic of our ion transport set up. (b) Water conductance of twelve salt-stabilized vermiculite membranes that were examined for their transport.

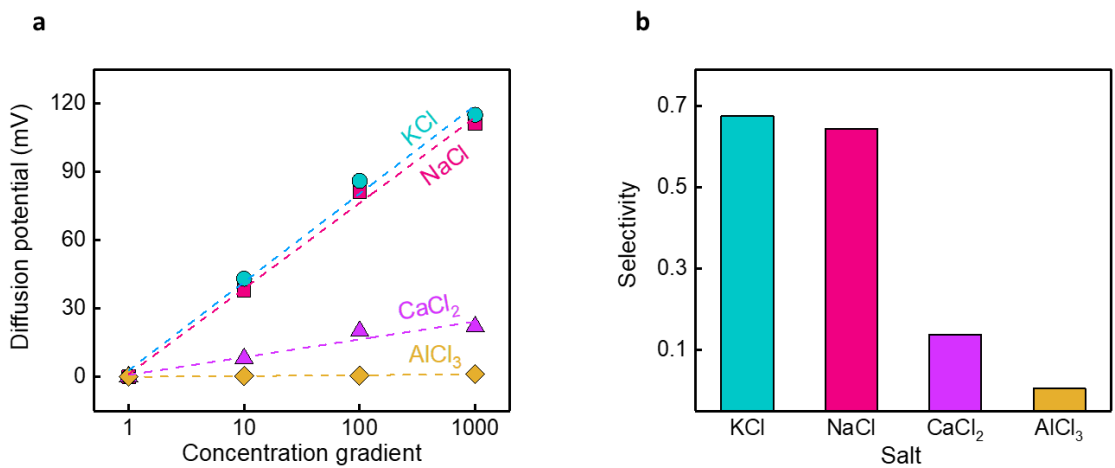

**Figure S11. Decreasing selectivity with increasing cation valence.** (a) Diffusion potential of KCl, NaCl, $CaCl_2$, and $AlCl_3$ through K-V, Na-V, Ca-V, and Al-V respectively at concentration gradients 10, 100, and 1000, with $C_H$ = 1 M and $C_L$ = 1 mM to 1 M. The diffusion potential is corrected for the contribution of redox and liquid junction potential. The latter arises from the intrinsic difference in the mobility of cations versus anions. (b) Selectivity of vermiculite membranes stabilized and studied with the same chloride solutions.

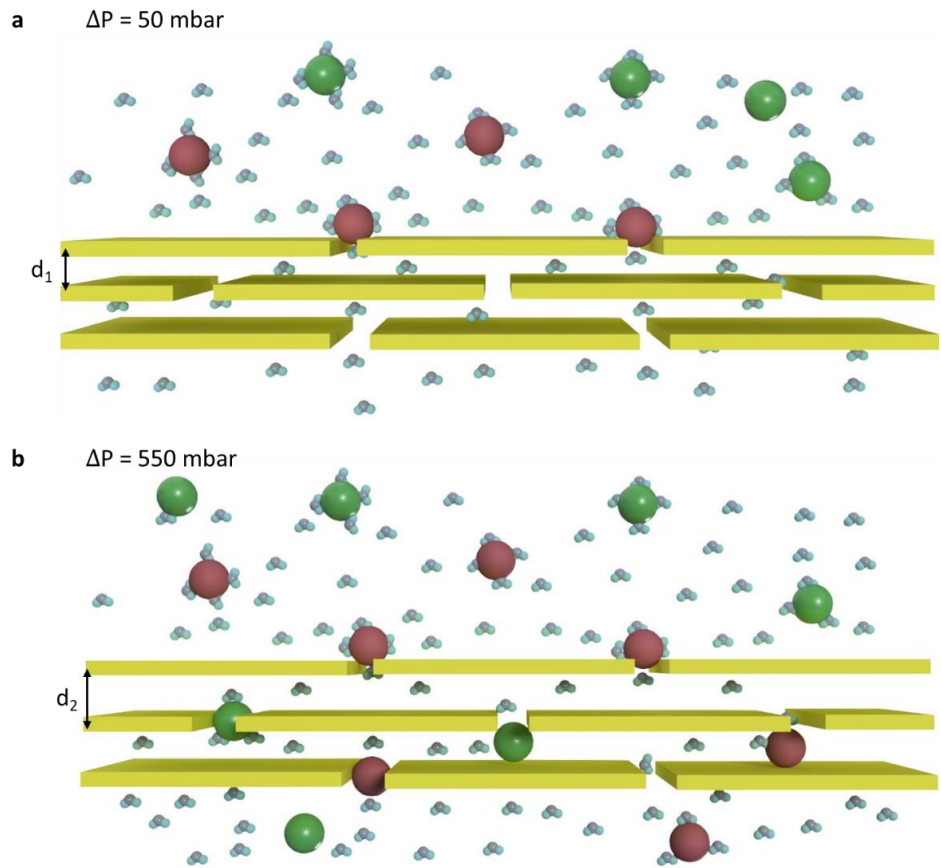

**Figure S12. Water and Ion flow.** Schematic of the water flow and NaCl ion rejection at pressure gradients of (a) 50 mbar and (b) 550 mbar. Here the red and green balls indicate Na and Cl ions, respectively. The ions are hydrated with water molecules surrounding the ion. The interlayer separation at 50 mbar is denoted as $d_1$ and at 550 mbar as $d_2$. The experimental result indicates that the interlayer separation $d_2 > d_1$, which leads to higher water flow and less rejection of salt ions at higher pressures.